\documentclass[reprint,aps,prl,amsmath,amssymb,amsfonts,showkeys,showpacs]{revtex4-1}

\usepackage{graphicx,bm,xcolor,hyperref,microtype}

\DeclareMathOperator{\Li}{Li}
\newcommand{\eps}{\epsilon}
\newcommand{\EHF}{\epsilon_{\rm HF}}
\newcommand{\Ec}{\epsilon_{\rm c}}

\newcommand{\mEh}{{\rm m}E_{\rm h}}
\newcommand{\mc}{\multicolumn}

\begin{document}

\title{Correlation energy of the one-dimensional Coulomb gas}

\author{Pierre-Fran\c{c}ois Loos}
\email{Corresponding author: loos@rsc.anu.edu.au}
\author{Peter M.W. Gill}
\email{peter.gill@anu.edu.au}
\affiliation{Research School of Chemistry, Australian National University, Canberra ACT 0200, Australia}


\begin{abstract}
We introduce a new paradigm for finite and infinite strict-one-dimensional uniform electron gases. In this model, $n$ electrons are confined to a ring and interact via a bare Coulomb operator. In the high-density limit (small-$r_s$, where $r_s$ is the Seitz radius), we find that the reduced correlation energy is $\Ec(r_s,n) =   \eps^{(2)}(n) + O(r_s)$, and we report explicit expressions for $\eps^{(2)}(n)$. In the thermodynamic (large-$n$) limit of this, we show that $\Ec(r_s) = - \pi^2/360 + O(r_s)$. In the low-density  (large-$r_s$) limit, the system forms a Wigner crystal and we find that $\Ec(r_s) = -[\ln(\sqrt{2\pi})-3/4]\,r_s^{-1} + 0.359933\,r_s^{-3/2} + O(r_s^{-2})$. Using these results, we propose a correlation functional that interpolates between the high- and low-density limits. The accuracy of the functional for intermediate densities is established by comparison with diffusion Monte Carlo results. Application to a non-uniform system is also reported. 
\end{abstract}

\maketitle

{\em Introduction.---}The usual paradigm for modeling uniform electron gases (UEG) in one dimension (1D) is the well-established Luttinger model \cite{Luttinger63}, which is exactly solvable by a Bogoliubov transformation technique \cite{Mattis65} and is particularly useful to study the low-energy spectrum of the 1D UEG \cite{Vignale}. This model has been motivated by the search for a replacement for the Fermi liquid theory in 1D which fails because of some divergences and makes conventional fermion many-body perturbation theory impracticable. The Luttinger model is shown to provide a general method for resumming all of the divergences \cite{Haldane81}. Although features associated with the Luttinger model have been observed in various systems, such as semiconductor quantum wires, ultracold atoms confined in elongated traps, GaAs quantum wells, carbon nanotubes, and many others \cite{Vignale}, the model is not strictly applicable to these systems because, whereas the standard Luttinger model assumes a short-ranged interaction, electrons actually interact via the long-range Coulomb force.

In the literature, the bare Coulomb operator $1/x$ (where $x$ is the interparticle distance) is usually avoided because of its divergence at $x=0$ and the intractability of its Fourier transform in 1D. Instead, most studies of the 1D UEG adopt a quasi-1D description by adding a transverse harmonic potential \cite{Pederiva02, Casula06, Lee11a} or use a potential of the form $1/\sqrt{x^2+\mu^2}$, where $\mu$ is a parameter that eliminates the singularity at $x=0$ while retaining the long-range Coulomb tail \cite{Schulz93, Fogler05a, Lee11a}. However, the introduction of the parameter $\mu$ is undesirable, for it modifies the physics of the system, especially in the high-density regime where neighboring electrons experience only the constant part of the potential. It is also unnecessary because, if the wave function vanishes when any two electrons touch, the Coulomb potential does not lead to energy divergence \cite{Astrakharchik11}. This allows us to apply the 1D Bose-Fermi mapping \cite{Girardeau60} which states that the ground state wave function of the bosonic (B) and fermionic (F) states are related by $\Psi_\text{B} (\bm{R}) = \left| \Psi_\text{F} (\bm{R})\right|$, where $\bm{R} = (\bm{r}_1,\bm{r}_2,\ldots,\bm{r}_n)$ are the one-particle coordinates. In case of bosons, the divergence of the Coulomb potential has the effect of mimicking the Pauli principle, which, for fermions, prohibits any two fermions from overlapping.  This implies that, for strictly 1D systems, the bosonic and fermionic ground states are degenerate and the system is ``spin-blind''. Consequently, the paramagnetic and ferromagnetic states are degenerate and we will consider only the latter \cite{Lee11a}.

{\em Model.---} In this Letter, we study a strictly 1D model inspired by the Calogero-Sutherland (CS) model \cite{Calogero69, Sutherland71b}. We consider the rotation-invariant  ground state of $n$ electrons on a ring of radius $R$. This yields a uniform electron density $\rho = n/(2\pi R) = 1/(2\,r_s)$, where $r_s = \pi R/n$ is the Seitz radius. This paradigm has been intensively studied as a model for quantum rings, both experimentally  \cite{Aronov93, Morpurgo98, Warburton00, Lorke00, Fuhrer01, Bayer03, Fuhrer04, Sigrist04} and theoretically \cite{Fogler05b, Emperador01, Pederiva02, Emperador03, Gylfadottir06, Aichinger06, Rasanen09, QuasiExact09, QR12} because of their rich electronic, magnetic and optical properties, such as the Aharonov-Bohm effect  and their potential application in quantum information theory. However, unlike the $1/x^2$ potential of the CS model, our electrons interact via the true Coulomb potential $1/x$.

Our work is inspired by three landmark papers.  First, following Gell-Mann and Brueckner \cite{GellMann57}, we study the reduced ({\it i.e.}~per electron) correlation energy, defined as the difference between the exact and Hartree-Fock (HF) reduced energies  $\Ec(r_s,n) = \eps(r_s,n) - \EHF(r_s,n)$, in the high-density (small-$r_s$) regime. We show that, despite using the true Coulomb operator, one can safely apply standard perturbation theory in this regime. Second, following Wigner \cite{Wigner34}, we study the correlation energy in the low-density regime, where the electrons crystallize to form a Wigner crystal. Combining the information obtained for these two limiting regimes, we propose a correlation functional that yields satisfactory estimates of the correlation energy for finite and infinite systems at high, intermediate and low densities. Third, following Ceperley and Alder \cite{Ceperley80}, we explore the accuracy of the correlation functional by comparing its predictions with accurate diffusion Monte Carlo (DMC) calculations on finite and infinite systems. We also report an application to a non-uniform system. Reduced energies and atomic units are used throughout.

{\em Hamiltonian.---} The Hamiltonian \footnote{We eschew the usual fictitious uniform positive background potential because its inclusion does not prevent a divergence of the Coulomb energy in 1D systems} of $n$ electrons on a ring of radius $R$ is
\begin{equation} \label{H}
	\hat{H} = - \frac{1}{2R^2} \sum_{i=1}^{n} \frac{\partial^2}{\partial\theta_i^2} + \sum_{i<j}^{n} \frac{1}{r_{ij}},
\end{equation}
where $\theta_i$ is the angle of electron $i$ around the ring center, and $r_{ij} = R \sqrt{2-2\cos\left(\theta_i - \theta_j\right)}$ is the interelectronic distance between electrons $i$ and $j$.

\begin{table*}
	\caption{
	\label{tab:Ec}
	Reduced correlation energy ($-\Ec(r_s, n)$ in millihartree) for $n$ electrons on a ring with various $r_s$. Subscripts represent the statistical errors in the last digits. The DMC results for $n\to\infty$ are taken from Ref.~\cite{Lee11a}.}
	\begin{ruledtabular}
	\begin{tabular}{cccccccccccccccc}		
		$r_s$	&	0		&	\mc{2}{c}{$0.1$}		&	\mc{2}{c}{$0.2$}			&	\mc{2}{c}{$0.5$}			&	\mc{2}{c}{$1$}				&	\mc{2}{c}{$5$}					&	\mc{2}{c}{$10$}					&	\mc{2}{c}{$20$}					\\
				\cline{2-2}		\cline{3-4}					\cline{5-6}						\cline{7-8}						\cline{9-10}					\cline{11-12}						\cline{13-14}						\cline{15-16}
		$n$		&	ISI		&	DMC		&	ISI		&	DMC			&	ISI		&	DMC			&	ISI		&	DMC			&	ISI		&	DMC			&	ISI			&	DMC			&	ISI			&	DMC			&	ISI			\\
		\hline			
		2		&	13.212	&	$13.0_0$	&	12.9		&	$12.77_0$	&	12.67	&	$12.15_0$	&	11.94	&	$11.250_0$	&	10.922	&	$7.111\,0_0$	&	6.733\,6  		&	$4.937\,7_0$	&	4.691\,9		&	$3.122\,02_0$	&	2.998\,22		\\
		3		&	18.484	&	$18.1_0$	&	18.0		&	$17.75_0$	&	 17.64	&	$16.76_0$	&	 16.53	&	$15.346_0$	&	14.997	&	$9.368\,6_0$	&	8.972\,2		&	$6.427\,0_0$	&	6.168\,8		&	$4.029\,51_0$	&	3.899\,54		\\
		4		&	21.174	&	$20.5_2$	&	20.6		&	$20.24_2$	&	 20.15	&	$19.00_1$	&	 18.82	&	$17.320_1$	&	17.003	&	$10.390\,2_1$	&	10.013\,0		&	$7.084\,5_1$	&	6.838\,2		&	$4.425\,31_1$	&	4.299\,35		\\
		5		&	22.756	&	$22.3_2$	&	22.2		&	$21.66_2$	&	 21.62	&	$20.33_1$	&	 20.15	&	$18.439_1$	&	18.156	&	$10.946\,3_1$	&	10.590\,2		&	$7.439\,0_1$	&	7.203\,4		&	$4.635\,56_1$	&	4.514\,64		\\
		6		&	23.775	&	$22.9_2$	&	23.2		&	$22.63_2$	&	 22.57	&	$21.14_1$	&	 21.00	&	$19.137_1$	&	18.887	&	$11.285\,1_1$	&	10.947\,3		&	$7.653\,4_1$	&	7.426\,8		&	$4.761\,89_1$	&	4.645\,07		\\
		7		&	24.476	&	$23.8_2$	&	23.8		&	$23.24_2$	&	 23.21	&	$21.70_1$	&	 21.58	&	$19.607_1$	&	19.384	&	$11.508\,5_1$	&	11.185\,3		&	$7.794\,8_1$	&	7.574\,5		&	$4.844\,00_1$	&	4.730\,64		\\
		8		&	24.981	&	$24.2_2$	&	24.3		&	$23.69_3$	&	 23.68	&	$22.11_1$	&	 22.00	&	$19.940_1$	&	19.739	&	$11.663\,8_1$	&	11.352\,9		&	$7.889\,8_1$	&	7.677\,7		&	$4.900\,68_1$	&	4.790\,10		\\
		9		&	25.360	&	$24.5_2$	&	24.7		&	$24.04_2$	&	 24.03	&	$22.39_1$	&	 22.31	&	$20.186_1$	&	20.002	&	$11.776\,5_1$	&	11.475\,8		&	$7.959\,5_1$	&	7.753\,0		&	$4.941\,36_1$	&	4.833\,27		\\
		10		&	25.651	&	$25.2_4$	&	24.9		&	$24.25_4$	&	 24.29	&	$22.62_1$	&	 22.55	&	$20.373_1$	&	20.204	&	$11.856\,8_1$	&	11.569\,0		&	$8.013\,4_1$	&	7.809\,8		&	$4.972\,54_1$	&	4.865\,69		\\
		\vdots	&	\vdots	&			&	\vdots	&				&	\vdots	&				&	\vdots	&				&	\vdots	&				&	\vdots		&				&	\vdots		&				&	\vdots		\\	
		$\infty$	&	27.416	&			&	26.6		&				&	25.90	&				&	23.96	&	$21.444\,1_2$	&	21.392	&	$12.317\,74_2$	&	12.091\,3		&	$8.292\,096_9$	&	8.120\,5		&	$5.132\,504_2$			&	5.039\,13		\\
	\end{tabular}
	\end{ruledtabular}
\end{table*}

{\em Hartree-Fock approximation.---}The HF wave function $\Psi_\text{HF}$ is a determinant of one-electron orbitals $\chi_k(\theta) = e^{i k \theta}$ with orbital energy $\kappa_k = k^2/(2R^2)$, where $k \in \mathbb{Z}$ if $n$ is odd and $k + \frac{1}{2} \in \mathbb{Z}$  if $n$ is even.  These orbitals form a Vandermonde matrix \cite{NISTbook} and, following the approach of Mitas \cite{Mitas06}, one discovers the remarkable result $\Psi_\text{HF}  \propto \prod_{i<j}^{n} \Hat{r}_{ij}$, where $\Hat{r}_{ij} = 2 R \sin[(\theta_i-\theta_j)/2]$ is a signed interelectronic distance. One sees that $\Psi_\text{HF}$ antisymmetric with respect to electron exchange and vanishes whenever $\theta_i = \theta_j$. The resulting HF reduced  energy is
\begin{equation} \label{EHF}
	\eps_\text{HF}(r_s,n) = \frac{n^2-1}{n^2} \frac{\pi^2}{24\,r_s^2} + \frac{1}{4\,r_s} \left( \sum_{k=1}^{n} \frac{4-1/n^2}{2k-1} - 3\right),
\end{equation}
where the first and second terms in \eqref{EHF} represent the kinetic and potential energies, respectively.  Although the latter is finite for finite $n$, it cannot be partitioned into Coulomb and exchange parts, because each diverges.
 
{\em High-density expansion.---} In the high-density ({\it i.e.}~small $r_s$) regime, the kinetic energy is dominant and it is natural to define a zeroth-order Hamiltonian and a perturbation by $\Hat{H}^{(0)} =  - 1/(2R^2) \sum_{i=1}^{n} \partial^2/\partial\theta_i^2$ and $\Hat{V} = \sum_{i<j}^{n} r_{ij}^{-1}$, and the reduced energy expansion is 
\begin{equation}
\label{expansion}
	\eps(r_s,n) = \frac{\eps^{(0)}(n)}{r_s^2} + \frac{\eps^{(1)}(n)}{r_s} + \eps^{(2)}(n) + O(r_s).
\end{equation}
The zeroth-order $\eps^{(0)}(n) = \langle \Psi_\text{HF} |\Hat{H}^{(0)}| \Psi_\text{HF} \rangle$ and first-order $\eps^{(1)}(n) = \langle \Psi_\text{HF}| \Hat{V}| \Psi_\text{HF} \rangle$ energies are found in \eqref{EHF}. The second-order energy is
\begin{equation} \label{MP2}
	\eps^{(2)}(n) = - \frac{1}{n} \sum_{ab}^{\text{occ}}\sum_{rs}^{\text{virt}} \frac{ |\langle \Psi_\text{HF} | \Hat{V} |  \Psi_{ab}^{rs} \rangle|^2}{\kappa_r + \kappa_s - \kappa_a - \kappa_b},
\end{equation}
where $\Psi_{ab}^{rs}$ is a doubly-substituted determinant  in which two electrons are promoted from $\{\chi_a,\chi_b\}$ to $\{\chi_r,\chi_s\}$.  If angular momentum is conserved, {\it i.e.}~$a+b=r+s$, then $\kappa_r + \kappa_s - \kappa_a - \kappa_b = (r-a)(r-b)/R^2$, and the Slater-Condon rules reveal that, if $a < b$,
\begin{equation}
	V_{a,b,r} = \langle \Psi_\text{HF} | \Hat{V} |  \Psi_{ab}^{rs} \rangle = \frac{2}{\pi} \sum_{k=r-b+1}^{r-a} \frac{1}{2k-1}.
\end{equation}
Combining these yields
\begin{equation} \label{MP2-final}
	\eps^{(2)}(n) = - \frac{1}{n} \sum_{a=-\frac{n-1}{2}}^{-1/2} \sum_{b=a+1}^{-a} \sum_{r=\frac{n+1}{2}}^\infty \frac{(2-\delta_{-a,b})V_{a,b,r}^2}{(r-a)(r-b)},
\end{equation}
which can be evaluated to give, for example,
\begin{align}
	\eps^{(2)}(2)	& = 1 - 10/\pi^2,				\\
	\eps^{(2)}(3)	& = 16/9 - 1436/(81\pi^2),	\\
	\eps^{(2)}(4)	& = 109/45 - 244168/(10125\pi^2).
\end{align}
Other values are shown in the second column of Table \ref{tab:Ec} and, in the $r_s \to 0$ limit, this gives all of the correlation energy. This many-electron system is one of the few for which one can obtain the \emph{exact} closed-form correlation energy for any value of $n$.

{\em Low-density expansion.---} In the low-density ($r_s \gtrsim 2$ \cite{Gylfadottir06}) regime, the electrons form a Wigner crystal. Using strong-coupling perturbation theory \cite{TEOAS09}, the energy can be written
\begin{equation}
\label{large-rs}
	\epsilon(r_s,n) = \frac{\eta^{(0)}(n)}{r_s} +    \frac{\eta^{(1)}(n)}{r_s^{3/2}}  + O(r_s^{-2}),
\end{equation}
where the first term represents the classical Coulomb energy of the crystal and the second is the zero-point energy of the electrons vibrating around their equilibrium positions.

The Wigner crystal consists of $n$ electrons separated by an angle $2\pi/n$ and is closely related to the one-dimensional Thomson problem. Thus, we have
\begin{equation} \label{EW}
	\eta^{(0)}(n) = \frac{\pi}{2n^2} \sum_{i<j}^n \csc\left[ \frac{(j-i)\pi}{n}\right].
\end{equation}

The second term in the expansion \eqref{large-rs} is found by summing the frequencies of the normal modes obtained by diagonalization of the Hessian matrix. For electrons on a ring, the Hessian is a circulant matrix and its eigenvalues and eigenvectors can be found in compact form, yielding 
\begin{equation} 
	\eta^{(1)}(n) =\frac{\pi^{3/2}}{4n^{5/2}} \sum_{i=1}^{n-1} \sqrt{\sum_{k=1}^{n-1} \frac{2-\sin^2\left(\frac{k \pi}{n}\right)}{\sin^3\left(\frac{k \pi}{n}\right)} \sin^2\left(i\frac{k \pi}{n}\right)}.
\end{equation}	

{\em Thermodynamic limit.---} In the $n\to\infty$ limit within the high-density regime, the kinetic energy
\begin{equation}
	\eps^{(0)} = \pi^2/24
\end{equation}
reduces to that of the ideal Fermi gas in 1D \cite{Vignale, Glomium11} and the slow decay of the Coulomb operator causes
\begin{equation}
	\eps^{(1)} = \ln\sqrt{n} + (\ln 2 + \gamma/2 - 3/4) + o(n^0)
\end{equation}
(where $\gamma$ is the Euler-Mascheroni constant \cite{NISTbook}) to grow logarithmically.

The limiting second-energy $\eps^{(2)} = - \pi^2/360$ can be found by converting the summations in \eqref{MP2-final} into integrals.  In this way, one finds
\begin{equation} \label{eq:Ec_hi}
	\eps_c(r_s) = - 0.027416 + O(r_s),
\end{equation}
and, in the dual thermodynamic/high-density limit, the exact correlation energy is therefore 27.4 millihartrees ($\mEh$) per electron.  Using a quasi-1D model with a transverse harmonic potential, Casula {\it et al.}~conclude that, in the high-density limit, the correlation energy vanishes \cite{Casula06}. This strikingly different prediction stresses the importance of employing a realistic Coulomb operator. 

In the $n\to\infty$ limit in the low-density regime, one finds
\begin{equation}
	\eta^{(0)} = \ln \sqrt{n} + \frac{\ln (2/\pi)+\gamma}{2} + o(n^0),
\end{equation}
which has the same logarithmic divergence as $\eps_\text{HF}$, but with a different constant term.  One can also show that
\begin{equation} 
	\eta^{(1)}  = \frac{1}{4\pi} \int_0^{\pi} \sqrt{2\Li_3(1) - \Li_3(e^{i \theta}) - \Li_3(e^{-i \theta})} d\theta,
\end{equation}	
where $\Li_3$ is the trilogarithm function \cite{NISTbook}. We have not been able to find this integral in closed form, but it can be computed numerically with high precision, and yields $\eta^{(1)} = 0.359933$, which is identical to the value found by Fogler in Ref.~\cite{Fogler05a} for an infinite ultrathin wire and a potential of the form $1/\sqrt{x^2+\mu^2}$. This shows that, unlike the high-density limit where the details of the interelectronic potential are critically important, the correct low-density result can be obtained by using a modified Coulomb potential. Thus, in the dual thermodynamic/low-density region, we have
\begin{equation} \label{eq:Ec_lo}
	\Ec(r_s) =  -[\ln(\sqrt{2\pi})-3/4]\,r_s^{-1} + 0.359933\,r_s^{-3/2} + O(r_s^{-2}).
\end{equation}
The same expansion can be derived for the infinite wire \cite{Fogler05a}, confirming the equivalence of the electrons-on-a-ring and electrons-on-a-wire models in the thermodynamic limit \cite{Glomium11}.

{\em Correlation functional.---} We now use the information obtained in the high- and low-density limits to build a correlation functional for finite and infinite 1D systems. We employ an interpolation between the high- and low-density limits inspired by the ``interaction-strength interpolation'' (ISI) expression of Seidl and coworkers \cite{Seidl00}. We define
\begin{multline}
\label{ISI}
	\Ec^{\text{ISI}}(r_s,n) = \frac{\eta^{(0)}(n)}{r_s}
	- \frac{[\eta^{(1)}(n)]^2}{2\epsilon^{(2)}(n)r_s^2}\Bigg[ \sqrt{1+ \alpha^2(n) r_s}  
	\\
	- \beta(n) \ln \left( \frac{\sqrt{1+ \alpha^2(n) r_s} + \beta(n)}{1+\beta(n)}\right) - 1\Bigg],
\end{multline}
with
\begin{gather}
	\alpha(n) = 2 \epsilon^{(2)}(n)\eta^{(1)}(n)/\eta^{(0)}(n)^2	\\
	\beta(n) = \epsilon^{(2)}(n)\eta^{(1)}(n)^2/\eta^{(0)}(n)^3-1
\end{gather}
Expression \eqref{ISI} reproduces the first term of the high-density expansion \eqref{eq:Ec_hi} and the first two terms of the low-density expansion \eqref{eq:Ec_lo}, thus giving correct energies at $r_s=0$ and vanishing at the correct rate for large $r_s$. 

{\em Discussion.---}In Table \ref{tab:Ec}, we report correlation energies for $2 \le n \le 10$ and various $r_s$ values from the high- and low-density regimes. Results obtained using the ISI method \eqref{ISI} are compared with DMC calculations. Our DMC code follows the implementation of Ref.~\cite{Reynolds82} and the energy at zero time-step is obtained by linear extrapolation. The extrapolated standard error is obtained by assuming that the data are Gaussian distributed \cite{Lee11b}.
The trial wave function is $\Psi_\text{T}= \Psi_\text{HF} \prod_{i<j}^n \left( \sum_{k=1}^{5} c_k r_{ij}^k \right)$ and the coefficients have been optimized using the procedure described in Ref. \cite{Umrigar05}.

For $0.1 \le r_s \le 0.5$, the ISI and DMC correlation energies agree to within 0.2 $\mEh$, which is remarkable given the simplicity of the functional.  The deviations increase to 0.3--0.4 $\mEh$ for $r_s=1$ and $r_s = 5$, but then decrease again to 0.1--0.2 $\mEh$ for $r_s = 10$ and $r_s = 20$. Overall, the ISI correlation functional gives reasonable estimates of the correlation energy for $2 \le n \le 10$. 

The functional can also be used to estimate the correlation energy in infinite systems and, in Table \ref{tab:Ec}, its predictions are compared with the DMC calculations of Lee and Drummond \cite{Lee11a} for an infinitely thin wire. We find that $\Ec^{\text{ISI}}$ underestimates the energies by 0.05, 0.23, 0.17 and 0.09 $\mEh$ for $r_s=$ 1, 5, 10 and 20, respectively, indicating that Eq.~\eqref{ISI} provides accurate correlation energy estimates in both finite and infinite uniform systems.

Moreover, the expression \eqref{ISI} can also be used as a correlation functional for non-uniform systems. For example, let us consider a 1D two-electron quantum dot (QD), {\it i.e.}~two electrons interacting via a bare Coulomb potential and confined by a harmonic potential of force constant $\omega^2$. Following Ref.~\cite{QR12}, we find that, for $\omega=1/2$, the exact reduced energy is $\epsilon^\text{QD} = 3/4$, the exact wave function is $\psi^\text{QD}(x_1,x_2) = (x_1-x_2)(1+|x_1-x_2|/2) \exp[-(x_1^2+x_2^2)/4]$ and the exact density is $\rho^\text{QD}(x) = \{7 + x (2 + x) [6 + x (2 + x)]\} \exp(-x^2/2)/(10\sqrt{2\pi})$, where $x_1$ and $x_2$ are the Cartesian coordinates of the electrons. Using a Gaussian expansion, we obtain $\EHF^\text{QD} = 0.7591$, which yields $\Ec^\text{QD} = -0.0092$. The expression \eqref{ISI} with $n=2$ yields
\begin{equation}
	\Ec^\text{QD}[\rho^\text{QD}] = \frac{1}{2} \int_{-\infty}^{\infty} \rho^\text{QD}(x) \Ec^\text{ISI}(\rho^\text{QD},2) dx = - 0.0106,
\end{equation}
which underestimates the reduced correlation energy by only 1.4 $\mEh$. We note that if one use the functional \eqref{ISI} with $n \to \infty$, the resulting correlation energy is found to be wrong by a factor two.

{\em Acknowledgements.---} We thank Neil Drummond for providing us the DMC correlation energies for $n\to\infty$ in Table \eqref{tab:Ec}, the NCI National Facility for a generous grant of supercomputer time and the Australian Research Council (Grants DP0984806, DP1094170, and DP120104740) for funding.

\end{document}